\documentclass[prd,twocolumn,showpacs]{revtex4}
\usepackage{graphicx}
\usepackage{latexsym}
\usepackage{amsmath}
\newcommand{\bq}{\begin{equation}}
\newcommand{\eq}{\end{equation}}
\newcommand{\bqn}{\begin{eqnarray}}
\newcommand{\eqn}{\end{eqnarray}}

\newcommand{\lb}{\label}
\begin{document}
\title{Thermodynamical properties of dark energy}
\author{Yungui Gong}
\email{yungui_gong@baylor.edu}
\affiliation{School of Physical Science and Technology,
Southwest University, Chongqing 400715, China}
\affiliation{CASPER, Department of Physics,
Baylor University, Waco, TX 76798, USA}
\author{Bin Wang}
\email{wangb@fudan.edu.cn}
\affiliation{Department of Physics, Fudan University,
Shanghai 200433, China}
\author{Anzhong Wang}
\email{anzhong_wang@baylor.edu}
\affiliation{CASPER, Department of Physics, Baylor University,
Waco, TX 76798, USA}
\begin{abstract}
We have investigated the thermodynamical properties of dark energy. Assuming
that the dark energy temperature $T\sim a^{-n}$ and considering that the volume of the
Universe enveloped by the
apparent horizon relates to the temperature, we have derived the dark energy entropy.
For dark energy
with constant equation of state $w>-1$ and the generalized Chaplygin gas,
the derived entropy can be positive
and satisfy the entropy bound. The total entropy,
including those of dark energy, the thermal radiation and the
apparent horizon, satisfies the generalized second law of thermodynamics.
However, for the phantom with constant equation of state, the positivity of entropy,
the entropy bound, and the generalized second law cannot be satisfied simultaneously.

\end{abstract}
\preprint{gr-qc/0611155}
\maketitle

Results from numerous and complementary observations show an emerging a paradigm `concordance cosmology'
indicating that our universe is spatially flat and composed of
about 70\% dark energy (DE) and about 25\% dark matter.  The weird DE is a major puzzle of physics now.
Its nature and origin have been
the intriguing subject of discussions in the past years. The DE has been sought within a wide range of physical phenomena,
including a cosmological constant, quintessence or an exotic field called phantom \cite{review}. Except the known fact that
DE has a negative pressure causing the acceleration of the universe, its nature still remains a complete mystery.
In the conceptual set up of the DE, one of the important
questions concerns its thermodynamical properties. It is expected that the thermodynamical consideration might shed some light
on the properties of DE and help us understand its nature.

The topic on the DE entropy, temperature and their evolution by
using the first law of thermodynamics was widely discussed in the
literature
\cite{others,odintsov,bousso05,santos,lima,pedro,phanthermo,wga,gong}.
It was found that the entropy of the phantom might be negative
\cite{lima,pedro,phanthermo}. The existence of negative entropy of the phantom could
be easily seen from the relation $Ts=\rho+p$ between the temperature
$T$, the entropy density $s$, the energy density $\rho$ and the
pressure $p$. Negative entropy is problematic if we accept that
the entropy is in association with the measure of the
number of microstates in statistical mechanics.  The intuition of
statistical mechanics requires that the entropy of all physical
components  to be positive. Besides if we consider the universe as a
thermodynamical system, the total entropy of the universe including
DE and dark matter should satisfy the second law of thermodynamics. The
generalized second law (GSL) for phantom and non-phantom DE has been
explored in \cite{phanthermo}. It was found that the GSL can be
protected in the universe with DE. The GSL of the universe with DE
has been investigated in \cite{wga,gong} as well. In order to rescue the
GSL of thermodynamics, Bekenstein conjectured that there exists an
upper bound on the entropy for a weakly self-gravitating physical
system \cite{bekenstein}. Bekenstein's entropy bound has received
independent supports \cite{wang}. A holographic entropy bound
\cite{hooft} was subsequently built and it was argued to be a real
conceptual change in our thinking about gravity \cite{witten}. The
idea of the holographic entropy bound was found to be a useful
tool in studying cosmology \cite{holography}.

In the discussion of thermodynamical properties of the universe, it
is usually assumed that the physical volume and temperature of the
universe are independent and by using the integrability condition
$\partial^2 S/\partial V\partial T=\partial^2 S/\partial T\partial
V$ and the first law of thermodynamics, one obtains the constant
co-moving entropy density. However, if we apply this treatment in
the universe with DE, we found some problems of the DE
thermodynamics \cite{gong}. Naively, we may think the DE temperature
is equal or proportional to the horizon temperature $T_H$. It was found
that the equation of state of the DE is uniquely determined and the
phantom entropy is negative \cite{gong}. Therefore, a general DE
model is not in thermal equilibrium with the Hawking radiation of
the horizon. Besides, although the GSL can be valid for $w>-1$, for
the phantom with $w<-1$, it was found that the GSL breaks down
due to the negative temperature deduced in the formalism where the
volume and the temperature are assumed to be independent
\cite{gong}. In summary, for the phantom, we either run into
negative entropy problem or the GSL is violated. It is more
realistic to consider that the physical volume and the temperature
of the universe are related, since in the general situation they both
depend on the scale factor $a(t)$. In the cosmological context, the
apparent horizon is important, since on the apparent horizon there
is the well-known correspondence between the first law of
thermodynamics and the Einstein equation \cite{cai}. On the other
hand, it was found that the apparent horizon is a good boundary for
keeping thermodynamical laws \cite{wga}. Considering the apparent
horizon as the physical boundary of the universe, it was found that
both the temperature and entropy can be positive for the DE,
including phantom. Furthermore, by considering the realistic case
that the physical volume and the temperature are related, the GSL is
proved to be always satisfied within the volume of the apparent
horizon \cite{gong}. Thus, in studying the DE thermodynamics, it is
more appropriate to consider the universe in which the volume and
the DE temperature are related.

In this work we will investigate the thermodynamical properties of
DE by assuming that the physical volume and the temperature are
not independent. Now again it is natural to think that the DE is in
thermal equilibrium with the Hawking radiation of the apparent
horizon. In this case, we found that the DE entropy is the dominant
entropy component and it becomes negative even for DE with
$w>-1$ \cite{gong}. Recall that the radiation temperature in the
universe scales as $T\sim a^{-1}$, so we assume here that the DE
temperature has a similar behavior $T\sim a^{-n}$ to avoid the
negative entropy problem, where $n$ is an arbitrary constant. It is
not necessary to take $n=1$ to ensure that the DE is in equilibrium
with the thermal radiation, since their dispersion relations could
be completely different \cite{lima,gong}. From the above
discussions, it is reasonable to expect that a physically acceptable
entropy of the DE should be positive and satisfy the entropy bound.
It should also satisfy the property required by the GSL.
Since the usual thermal radiation temperature in the
universe decreases as the universe expands, we expect that the DE
temperature also preserves this property.

By using the first law of thermodynamics $TdS=dE+pdV$ for the DE,
and considering the volume of the universe within the apparent horizon $V=4\pi \tilde{r}_A^3/3$,
the total DE $E=\rho V$,
we can express the DE entropy as \cite{gong}
\bq
\lb{theq2}
TdS=-\frac{2\pi}{3}\left(\frac{8\pi
G}{3}\right)^{-3/2}\rho_t^{-5/2} (\rho_t+3p_t)d\rho,
\eq
where the Friedmann equation and the energy conservation law have been used in the derivation,
$\rho_t$ and $p_t$ denote the total energy density and pressure, respectively.
Taking derivative with respect to time on both sides of the above equation, we have
\bq
\lb{sdot} \dot{S}=2\pi\left(\frac{8\pi
G}{3}\right)^{-3/2}\rho_t^{-5/2} (\rho_t+3p_t)H(\rho+p)/T.
\eq
It can be seen that $\dot{S}\ge 0$ ($<0$) if $(\rho+p)/T\ge 0$ ($<0$)
during radiation dominated era (RD) and matter dominated era (MD), and $\dot{S}\le 0$ ($>0$) if $(\rho+p)/T\ge 0$
($<0$) during DE domination. In the phantom domination, the apparent horizon entropy decreases
as the Universe expands \cite{gong}. This requires $\dot{S}>0$ and $|1+3w|T_H/2>T>0$ to protect
the GSL. Thus in the phantom domination era, the
temperature of the phantom has to be positive to rescue the GSL.

The radiation entropy can be obtained as usual $S_r=s V$, where $s=\sigma/a^3$ is the physical
entropy density and $\sigma$ is the constant co-moving
entropy. For DE with constant equation of state $w$,
using the Friedmann equation, the entropies of the radiation and the apparent horizon are
\bq
\lb{sreq}
S_r=S_{r0}\,x^{-3}\,\Omega_t(x)^{-3/2},
\eq
\bq
\lb{saeq}
S_A=S_{A0}\Omega_t(x)^{-1},
\eq
where $\Omega_t(x)=\Omega_{m0}x^{-3}+\Omega_{r0}x^{-4}+\Omega_{w0}x^{-3(1+w)}$
and $x=a/a_0$.

To get the DE entropy we need to solve Eq. (\ref{theq2}) by assuming $T=T_0(a/a_0)^{-n}$.
In the evolution of the universe, the solution to Eq. (\ref{theq2}) is given in the form
\bq
\lb{swaeq}
\frac{S_w}{S_{r0}}\frac{\Omega_{r0}T_{w0}}{\Omega_{w0}T_{r0}}=\begin{cases}
\frac{9(1+w)}{4(n+3(1-w))}\Omega_{r0}^{-3/2}x^{n+3(1-w)},\\
S_{w1}+\frac{9(1+w)}{8(n+3/2-3w)}\Omega_{m0}^{-3/2}x^{n+3/2-3w},\\
S_{w2}+\frac{9(1+w)(1+3w)}{8(n+3(1+w)/2)}\Omega_{w0}^{-3/2}x^{n+3(1+w)/2},
\end{cases}
\eq
for the RD, MD and DE domination respectively, where $S_{w1}$ and $S_{w2}$ are integration constants.

As we mentioned previously, the intuition of the statistical mechanics
requires positive entropy. We expect that
this should also hold for the entropy of DE if it
is supposed to keep the same microscopic meaning. From Eq. (\ref{swaeq}) we
learn that for DE with constant equation of state $w>-1$, non-negative $S_w$ can be obtained if $n>3w-3$ during RD.
During DE domination, if $n>-3(1+w)/2$, then $S_w\rightarrow -\infty$ when
$a\rightarrow \infty$. Thus, to get positive entropy for the DE, the parameter $n$ should be chosen within the range
$-6<3w-3<n<-3(1+w)/2<0$.

\begin{figure}
\centering
\includegraphics[width=8cm]{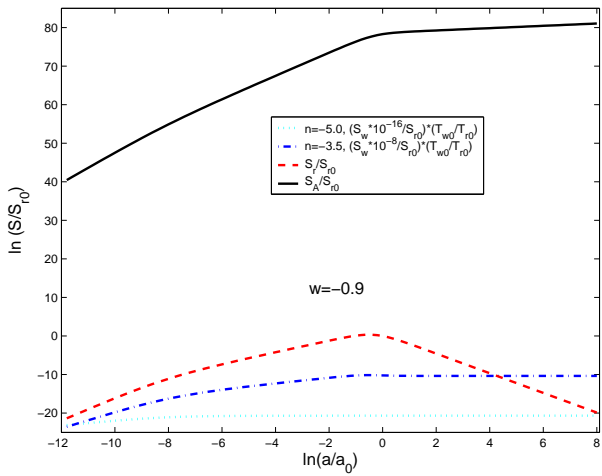}
\caption{The evolution of $S_w$ for $w=-0.9$ and $S_r$. The dotted line is for
$(S_w/S_{r0})\times(T_{w0}\times 10^{-16}/T_{r0})$ with $n=-5.0$, the dash-dot
 line is for $(S_w/S_{r0})\times(T_{w0}\times 10^{-8}/T_{r0})$
with $n=-3.5$, the dash line is for $S_r/S_{r0}$, and the solid line is for
the apparent horizon entropy $S_A/S_{r0}$.}
\label{fig1}
\end{figure}

This parameter range of $n$ can be further constrained if we express the solution of Eq. (\ref{theq2}) as
\bq
\lb{swsol}
\begin{split}
\frac{S_w}{S_{r0}}=&\frac{3}{4}(1+w)\frac{\Omega_{w0}}{\Omega_{r0}}
\frac{T_{r0}}{T_{w0}}\left[x^{-3+n-3w}\Omega_t(x)^{-3/2}\right.\\
&\left.-(n-3w)\int_0^{a/a_0} x^{-4+n-3w}\Omega_t(x)^{-3/2}dx\right].
\end{split}
\eq
If $n>3w$, the second term in the above equation is negative, which might lead $S_w$
to be negative. Therefore, we need to restrict $3w-3<n<3w$ to
ensure the positivity of $S_w$.
Note that for radiation, $n=3w=1$, Eq. (\ref{swsol}) reduces to Eq. (\ref{sreq}).
Since $n<0$, the
dark energy temperature will increase with the scale factor $a$
and at the present moment $T_{w0}\gg T_{r0}$.
During RD and MD, it can be seen that both
 DE entropy and the radiation entropy increase. However, if one notes that
$\dot{S}_w/\dot{S}_r=3(1+w)\rho_w T_r/(4\rho_r T_w)<1$, the DE
entropy increases slower than the radiation entropy. During the DE domination,
both the DE entropy and the radiation entropy decrease, $S_w\rightarrow
S_{w2}>0$ and $S_r\rightarrow 0$ when $a\rightarrow \infty$, so
$S_w>S_r$ in the future.
Since the apparent horizon entropy increases during the DE domination,
$\dot{S}_A=3(1+w)S_AH$, $\dot{S}_w=(n+3(1+w)/2)S_w H$ and
$\dot{S}_r=3(1+3w)S_r H/2$,
so the GSL is always respected for the DE with constant equation of state $w>-1$. To see these points
more clearly, we solve Eq. (\ref{sdot}) numerically by choosing
$w=-0.9$, $\Omega_{m0}=0.3$, $\Omega_{w0}=0.7$ and
$\Omega_{r0}=8.35\times 10^{-5}$. The results for $n=-5.0$ and $n=-3.5$
are shown in Fig. \ref{fig1}. The numerical results confirm that by
constraining $3w-3<n<3w$, $S_w$ is positive. It is easy to see
that the DE entropy and the radiation
entropy are much smaller compared to the apparent horizon entropy,
thus the entropy bound is always held.
Although the radiation entropy and the DE entropy may decrease
in the DE domination, due to their very small scale, their decreasing
behaviors can be overcome by the increase of the entropy on the apparent horizon. Thus,
including the total entropy in the
universe and the entropy of the apparent horizon,
we find that the GSL is protected.

Now we come to consider the phantom with constant equation of state $w<-1$.
In the RD and MD eras,
if $n<3w-3$, $S_w$ is positive but it decreases starting from $\infty$ as the universe expands.
The entropy bound is violated at very early times.
If $n>-3(1+w)/2$, $S_w$ is negative during RD and MD and $S_w\rightarrow\infty$
when $a\rightarrow \infty$, so in the future although the GSL can be protected, the entropy bound will be violated.
If $3w-3<n<-3(1+w)/2$, $S_w$ is negative.
Thus, for the phantom with constant equation of state, it seems impossible to
get a viable thermodynamics. The requirements of the positivity of DE entropy, the entropy bound and the GSL
cannot be met  simultaneously.
In \cite{phanthermo}, the authors used the future event horizon
to study the phantom thermodynamics and found that the GSL could be respected
if the phantom entropy is negative.
The problem with the future event horizon is that for
the universe with DE with equation of state $w\neq -1$, the thermodynamical
description breaks down on the event horizon \cite{wga}.
Furthermore, the definitions of the event horizon temperature and entropy could be
less certain than a guess. Even if we use the similar temperature and entropy
definitions of the apparent horizon for the future event horizon, the first law
of the thermodynamics was not satisfied \cite{wga}.

\begin{figure}[htp]
\centering
\includegraphics[width=8cm]{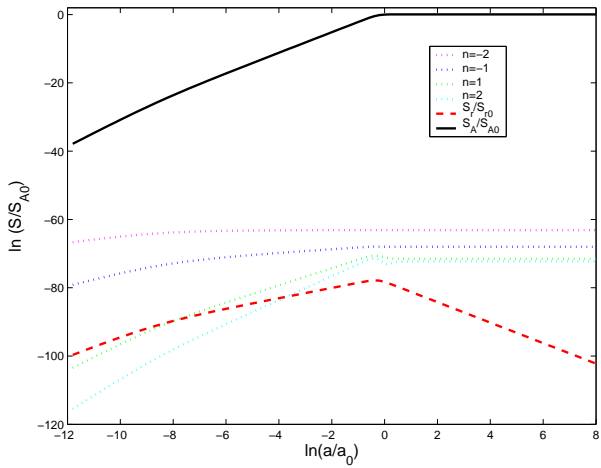}
\caption{The evolutions of $S_c$, $S_r$ and $S_A$ with $w_{c0}=-0.88$ and $\alpha=1.57$.
The dotted lines are for $(S_c/S_{A0})\times(T_{c0}/T_{r0})$,
from top to down are $n=-2$, $n=-1$,
$n=1$ and $n=2$ (Note that $S_c$ is negative after MD when $n=2$).  The
dash line is for $S_r/S_{A0}$, and the solid line is for $S_A/S_{A0}$.}
\label{fig3}
\end{figure}

In the above discussion we have concentrated ourselves
on the DE with constant equation of state.
To study the thermodynamics of a dynamic DE, we will use
the generalized Chaplygin gas (GCG) \cite{chap} as an example.
When the universe is dominated by the GCG, the entropies of the apparent
horizon and the radiation read \cite{gong}
\bq
\lb{sachap1}
S_A=S_{A0}\,\Omega_c^{-1},
\eq
\bq
\lb{chaprs1}
S_r=S_{r0}\left(\frac{a}{a_0}\right)^{-3}\Omega_c^{-3/2},
\eq
where $\Omega_c=[-w_{c0}+(1+w_{c0})(a/a_0)^{-3(1+\alpha)}]^{1/(1+\alpha)}$.
The entropy for the GCG can be obtained by solving Eq. (\ref{theq2}), which can be expressed as
\begin{widetext}
\bq
\lb{chapsc}
\frac{S_c}{S_{r0}}\frac{T_{c0}}{T_{r0}}
=\begin{cases}
\frac{9}{4(n+3)}(1+w_{c0})^{1/(1+\alpha)}\Omega_{r0}^{-5/2}x^{n+3}, & {\rm RD},\\
S_{c1}+\frac{9}{8(n+3/2)}(1+w_{c0})^{-1/2(1+\alpha)}\Omega_{r0}^{-1}
x^{n+3/2}, & a\ll a_0,\\
S_{c2}-\frac{9}{4(n-3-3\alpha)}(1+w_{c0})(-w_{c0})^{-1-1/2(1+\alpha)}
\Omega_{r0}^{-1}x^{n-3(1+\alpha)}, & a\gg a_0,
\end{cases}
\eq
\end{widetext}
where $S_{c1}$ and $S_{c2}$ are integration constants.
To have $S_c\ge 0$, the parameter $n$ must satisfy the condition $-3<n<3(1+\alpha)$.
Numerical results show that this condition is not enough. For example, if we choose
$w_{c0}=-0.88$ and $\alpha=1.57$, which are the best fitting values from observations \cite{chap}, we find that
$S_w$ is negative after MD when $n=2$.
At late times, $a\rightarrow \infty$,
$S_c\rightarrow S_{c2}$. For positive entropy, $S_c$ will be greater than
$S_r$ at late times since $S_r\rightarrow 0$.
The range of $n$ to keep $S_c$ positive can be more confined by numerical calculation.
Choosing appropriate $n$ to ensure $S_c$ to be positive, we have shown the numerical
results in Fig. \ref{fig3} on the evolution of entropies of GCG, radiation and the apparent horizon.
When $n<0$, $T_c$ increases with
the expansion of the universe and the numerical results show that $S_c$ can be less than $S_r$
during RD and MD eras if $T_{c0}/T_{r0}$ is large enough. If $n>0$, then $T_c$
decreases as the universe expands and $S_c$ increases
faster than $S_r$ during RD and MD eras.
When $n=1$, the GCG and the radiation temperatures evolve in the same way
and $S_c$ can be larger than $S_r$ during MD
as shown in Fig. \ref{fig3}. It is clear from Fig. \ref{fig3} that
compared to the apparent horizon entropy, $S_r$ and $S_c$
are negligible, thus the entropy bound can be protected for the GCG case.
In addition, the GSL can also be saved in the GCG case,
since the total entropy evolves basically in the same way as
the entropy of the apparent horizon. Though in the GCG dominated period,
$S_r$ decreases as the universe expands, owing to its
negligible value compared to the apparent horizon entropy, its decrease
can be overcome by the increase of the apparent horizon.

In summary, in this work we have investigated the thermodynamical
properties of the DE. In calculating the DE entropy we have
considered the volume of the universe enveloped by the
apparent horizon and assumed that the physical
volume and the temperature are related.  The apparent horizon is a good
boundary for studying cosmology, since on the apparent
horizon there is the well-known correspondence between the first
law of thermodynamics and the Einstein equation \cite{cai}.
Furthermore, it has been found that the apparent horizon
is good in keeping thermodynamical laws \cite{wga}.
Assuming that the temperature of the DE has the form $T\sim a^{-n}$,
we have derived the evolution of the DE entropy.
For the DE with constant equation of state $w>-1$, we have
found the appropriate range of $n$ for keeping DE entropy to be positive,
which is the requirement of the statistical understanding of the concept of entropy.
In this range of $n$, the entropy bound and the GSL can also be protected.
The negative point is that the allowed range of $n$ for giving physically
acceptable DE entropy leads
the DE temperature to increase as the universe expands, which is
different from the behavior of the thermal temperature that decreases
along the expansion of the universe. This conflict could be overlooked
since the DE temperature and the thermal temperature may have different dispersion relations
\cite{lima,gong},
and it is not necessary that these two different temperatures behave accordingly.
In the era of phantom domination, the GSL requires that the phantom entropy increases as
the universe expands and the
phantom temperature $T$ satisfies the condition $|1+3w|T_H/2>T>0$. Since the
horizon entropy decreases to zero as the universe expands, the holographic entropy bound
will be violated if the phantom entropy is positive.
For the phantom with constant equation of state $w<-1$, we found that there is no common
range of $n$ so that the positivity of the entropy, the entropy bound and the GSL
can all be satisfied. The physical requirement on the DE entropy does not favor
the phantom with constant equation of state.
We have also extended our investigation to the dynamical DE by using the GCG as an example.
We have found that by appropriately choosing parameters,
we can have positive DE entropy, and meanwhile we can
protect the holographic entropy bound and the GSL.
Within the allowed parameter range for physically acceptable DE entropy,
the DE temperature can decrease
and it can even scale in the same way as
the radiation temperature does as the universe expands.

\begin{acknowledgments}
Y.G.G. is supported by Baylor University, NNSFC under Grants No. 10447008
and No. 10605042, CMEC under Grant No. KJ060502, and SRF for ROCS, State
Education Ministry. The work of B.W. was partially supported by NNSF of
China, the Ministry of Education of China, and the Shanghai Education
Commission.
\end{acknowledgments}

\end{document}